\documentclass[journal,12pt,onecolumn,draftclsnofoot]{IEEEtran}
\usepackage{amsmath,amsfonts}
\usepackage{algorithmic}
\usepackage{algorithm}
\usepackage{array}
\usepackage[caption=false,font=normalsize,labelfont=sf,textfont=sf]{subfig}
\usepackage{textcomp}
\usepackage{stfloats}
\usepackage{url}
\usepackage{verbatim}
\usepackage{graphicx}
\usepackage{ntheorem}
\usepackage[algo2e]{algorithm2e} 
\usepackage{amsmath}
\usepackage{cite}

\usepackage{amsfonts}
\usepackage{lipsum}
\usepackage{mathtools}
\usepackage{cuted}
\usepackage{setspace}
\usepackage{multicol}

\usepackage{booktabs, multirow}
\usepackage{siunitx}

\usepackage{kantlipsum}

\usepackage{float}

\makeatletter

\newcommand{\Rmnum}[1]{\expandafter\@slowromancap\romannumeral #1@}

\begin{document}

\title{Distributed Resource Allocation for D2D Multicast in Underlay Cellular Networks}

\author{\IEEEauthorblockN{Mohd Saif Ali Khan*, Ajay Bhardwaj$^+$, and Samar Agnihotri*}\\
\IEEEauthorblockA{*Indian Institute of Technology Mandi, HP, India} \\
\IEEEauthorblockA{$^+$SRM University, Amaravati, AP, India}\\
Email: saifalikhan00100@gmail.com, ajaybhardwaj99@gmail.com, samar.agnihotri@gmail.com%
}




\maketitle

\begin{abstract}
We address the problem of distributed resource allocation for multicast communication in device-to-device (D2D) enabled underlay cellular networks. The optimal resource allocation is crucial for maximizing the performance of such networks, which are limited by the severe co-channel interference between cellular users (CU) and D2D multicast groups. However, finding such optimal allocation for networks with large number of CUs and D2D users is challenging. Therefore, we propose a pragmatic scheme that allocates resources distributively, reducing signaling overhead and improving network scalability. Numerical simulations establish the efficacy of the proposed solution in improving the overall system throughout, compared to various existing schemes.
\end{abstract}


\section{INTRODUCTION}
\IEEEPARstart{T}{he} rapid growth of mobile devices and data-intensive applications running on them has resulted in a dramatic increase in data requirements~\cite{cisco2020cisco}. Trying to address this challenge has become a primary concern for both the telecommunication services sector and the academic community. In the recent years, underlay D2D communication has emerged as a viable solution to address this challenge. It enables proximate mobile users to communicate with one another using cellular resources while bypassing the base station (BS)~\cite{feng2013device,liu2014device}. Due to its short-range exchange of information, D2D communication provides several benefits in cellular networks, such as increased throughput, lower latency, and enhanced spectral and energy efficiencies~\cite{bhardwaj2018energy,zeng2022joint,meshgi2017optimal}. By capitalizing on these advantages, D2D communication has the potential to revolutionize wireless communication and support mobile users' ever-increasing data needs. 

D2D-multicast communication has several advantages over D2D unicast or BS-based multicast communication, especially for the applications like weather forecasting and location-based advertisements that involve the same data to be distributed to nearby users~\cite{bhardwaj2019multicast}. This may allow more efficient network resource utilization and may mitigate severe bottlenecks in data-centric cellular networks. Furthermore, D2D-multicast facilitates and improves direct, backhaul-free communication among devices with similar content requirements, which can substantially reduce network congestion while enhancing data transfer rates. In conclusion, D2D multicast communication seems a promising option for improving wireless communication and meeting the ever-increasing requirements of data-intensive applications. Though underlay D2D multicast communication holds a great deal of promise, deploying it in cellular networks comes with several inherent challenges.

The interference that underlay D2D multicast communication may cause to primary cellular users poses one of the most significant challenges. Furthermore, D2D multicast nodes' battery power may quickly deplete when dealing with interference and sending information, leading to decreased network capacity and reliability. These issues in underlay D2D multicast communication must be addressed in order to be successfully deployed and widely adopted. 

To mitigate the co-channel interference between CUs and D2D multicast groups (MGs), numerous resource allocation techniques have been proposed in the literature~\cite{meshgi2017optimal,zhang2018resource,kim2017resource,bhardwaj2017interference,zhao2017social}. However, most of them are centralized in nature, thus involving significant signaling overhead and delays as the BS computes a solution to resource allocation problem, making it unsuitable for large-scale networks. To address this issue, distributed schemes for resource allocation have been proposed recently~\cite{hmila2020distributed,elnourani2021distributed}. In~\cite{hmila2020distributed}, the authors have proposed semi-distributed solution by using coalition game framework and fractional programming. In~\cite{elnourani2021distributed}, the authors have used fractional programming approach to tackle the joint power and channel allocation. However, it is important to note that this approach has a significant signaling overhead, due to message passing between the BS and MGs. 

The objective of our study is to solve the resource allocation problem in a distributed manner. We propose a novel two-step strategy to achieve this. In the first step, by utilizing the notion of distributed graph coloring, we present a distributed channel allocation scheme that allows cellular users to allocate channels distributively without incurring significant signaling overhead or requiring centralized computation. In the second step, power allocation among the D2D multicast groups, over a particular channel, is carried out using a distributed power allocation scheme to minimize the interference at the CUs while ensuring efficient resource utilization. With this, we achieve an efficient and scalable alternative to centralized resource allocation without compromising on the performance achievable by the centralized scheme, as confirmed by our numerical evaluation.

\textit{Organization:} The work is organized as follows: Section \ref{system_model} explains the system model used in this paper. Section \ref{channel_allocation} proposes a distributed channel allocation approach, while Section \ref{power_allocation} presents a distributed power allocation scheme that complements the proposed channel allocation approach. Section \ref{simulation} evaluates the performance of our proposed approach. Finally, Section \ref{conclusion} concludes the paper.

\section{SYSTEM MODEL}
\label{system_model}
In our proposed system model, we consider the coexistence of CUs and D2D multicast group transmitters (MGTx) in a two-dimensional space $\mathbb{R}\textsuperscript{2}$. We model these two types of users as two independent and stationary homogeneous Poisson point processes (PPP), and having the densities $\lambda{\textsubscript{c,u}}$  and $\lambda{\textsubscript{g,t}}$, respectively. Using the properties of homogeneous PPP, we calculate the number of mobile users as follows: 
\begin{align}
\label{deqn_ex1a}
Pr[\textit{N}(\textit{S})=k] = \frac{(\lambda S)^{k}}{k!}e^{-\lambda \textit{S}}
\end{align}
As the process is homogeneous, the $\lambda$ is constant and denotes the average node density per unit area. To make the system more pragmatic, we assume that the D2D multicast networks follow a Mat{\'e}rn cluster process~\cite{liu2013transmission}, where several clusters co-exist in $\mathbb{R}\textsuperscript{2}$, and each cluster center (MGTx) is encircled by a homogeneous and independent child PPP of radius \textit{d\textsubscript{r}} and having a density of $\lambda{\textsubscript{g,r}}$ modeled as the D2D receivers. The radius \textit{d\textsubscript{r}} of MGs in D2D multicast is determined by the maximum power that MGTx can transmit and signal detection threshold, which ensures that all MG receivers (MGRx) within this radius can receive the transmitted data with sufficient quality for reliable communication. This configuration results in each MG having a different number of MGRx. The parameter $ \mathcal{U}\textsubscript{g} $ represents the set of MGRx of the $g^\text{th}$ MG. The sets $\mathcal{C}$ and $\mathcal{G}$ represent the set of all CUs and MGs repectively.
To maximize the spectral efficiency, we assume that a single CU uplink channel is shared by multiple MGs; thus, the signal-to-interference-plus-noise ratio achieved by a CU and D2D MG are given respectively by:
\begin{align}
\Gamma^{{k}}_{{c}} &= \frac{p\textsubscript{c,k}h^{k}_{c,b}}{\sum\limits^{\mathcal{G}}_{g=1}{a_{g,k}}{h^{k}_{g,b}p\textsubscript{g,k}} + N\textsubscript{0}}, \label{eq_cu_sinr}\\
\Gamma^{{k}}_{{r,g}} &= \frac{p\textsubscript{g,k}h^{k}_{g,r}}{\sum\limits^{\mathcal{G}}_{j=1,j \neq g}{a_{g,k}}{h^{k}_{j,r}p\textsubscript{j,k}} + h^{k}_{c,r}p\textsubscript{c,k}+ N\textsubscript{0}}, \label{eq_d2d_sinr}
\end{align}
where $a_{g,k}$ is the binary variable that indicates whether the $g^{\textrm{th}}$ MG shares the resources with the $k^\textrm{th}$ CU or not; $h^{k}_{g,r}$ and $h^{k}_{c,b}$ are the channel gains between $g^\text{th}$ MGTx and $r^{th}$ MGRx, and the $k^\text{th}$ CU and the BS, respectively, that incorporates both small scale fading as well as large scale fading; $p_\text{c,k}$ and $p_\text{g,k}$ denote the transmit powers of the $k^\text{th}$ CU and the $g^\text{th}$ MG, respectively; $N_0$ denotes the variance of the AWGN noise, and $B$ represents the bandwidth of each channel. Let ${R}^{k}_{c}$ be the data rate of the CU that transmits on the $k^{th}$ channel and ${R}^{k}_{g}$ be the rate that can be attained by the $g^{th}$ MG by considering the SINR of the worst MGRx. These rates are given by the following equations, respectively. 
\begin{align}
{R}^{k}_{c} &= B\log_{2}(1+\Gamma^{k}_{c}), \label{eq_cu_rate}\\
{R}^{k}_{g} &={\left | \mathcal{U}\textsubscript{g} \right |B\log_{2}\left(1+ \min_{r\in\mathcal{U}\textsubscript{g}} \Gamma^{{k}}_{{r,g}}\right)}, \label{eq_d2d_rate}
\end{align}

The maximum transmit powers of the CUs and MGTXs that can be allocated are $P_c^{max}$ and $P_g^{max}$ respectively. Using \eqref{eq_cu_rate} and \eqref{eq_d2d_rate}, we formulate an optimization problem with total sum-throughput as a objective function as:
\begin{align}
\mathbf{P1:}~~~ &\underset{{a_{g,k}, p_{g,k}}} {\max~}
\left( \sum\limits_{c \in  \mathcal{C} } R_{c}^{k} + \sum\limits_{g \in \mathbb{G}_{k}} R_{g}^k \right) \label{main_problem} \\
\text{s.t.} \quad 
& \text{C}_1: 0 \leq p_{g,k} \leq P_{g}^{max}, \forall k \in \mathcal{C},  \nonumber\\ 
& \text{C}_2: \Gamma_{r, g}^k \geq \Gamma_{g,k}^{\textrm{th}}, \Gamma_c^{k} \geq \Gamma_{c,k}^{\textrm{th}}, \nonumber\\
& \text{C}_3: {\sum_{k \in \mathcal{C}} a_{g,k} =1}, g \in \mathcal{G}, k \in \mathcal{C}, \nonumber 
\end{align}
where $\text{C}_1$ limits the maximum power allocated to D2D MGTXs, $\text{C}_2$ guarantees minimum achievable rate for every CU and receivers of every MG, and $\text{C}_3$ guarantees that each MG can be assign only one channel.

Problem $\mathbf{P1}$ is an instance of Mixed-Integer Nonlinear Programming (MINLP) problem, which are NP-hard, in general. To reduce the computational costs associated with solving this problem, a pragmatic approach is used that splits the original problem into two sub-problems that are solved sequentially, namely the channel allocation problem and the power allocation problem. The correlation between these sub-problems is that power allocation is dependent on the channels assigned in the channel allocation step. The iterative optimization method ensures that the two sub-problems converge, resulting in a set of sub-optimal solutions that satisfy all constraints of the original problem $\mathbf{P1}$ jointly. While the solutions obtained in each sub-problem may not be globally optimal due to the inherent NP-hardness of MINLP problems, their combination results in a solution for the original problem $\mathbf{P1}$ that meets the objectives and constraints.
This work proposes distributed schemes to address these sub-problems.

\section{Distributed Channel Allocation}
\label{channel_allocation}
The problem of distributed channel allocation is formulated as distributed graph coloring problem. The channels are assigned to various MGs using distributed graph coloring~\cite{szegedy1993locality}. If number of MGs is greater than the number of CUs, then the algorithm utilizes all channels. Let $N_p$ and $N_c$ be number of MGTx and CUs, respectively. The achievable rate of the $k^\textrm{th}$ CU when no MG is sharing the channel is given by:
\begin{align}
{R}^{k} &= B\log_{2}\left(1+\frac{p\textsubscript{c,k}h^{k}_{c,b}}{ N\textsubscript{0}}\right)\label{eq_8}
\end{align}

Using (\ref{eq_cu_rate}), (\ref{eq_d2d_rate}) and (\ref{eq_8}), the overall change in sum-throughput of the system due to the $g^{th}$ MG, when  $a_{g,k}=1 \ \forall g \in \mathcal{G}$, is given as follows:
\begin{align}
   \Delta{R^{k}_{g}} = R^{k}_{g} + R^{k}_{c} - {R}^{k}\label{9}
\end{align}

Let $ \mathcal{C}_{g} \subset \mathcal{C} $ be the set of available channels for the $g^{th}$ MGTx. If $\Delta{R^{k}_{g}} > 0$ then $g^\text{th}$ MG can share the $k^\text{th}$ CUs channel and then $\mathcal{C}_{g}$ will be updated as $\mathcal{C}_{g} = \mathcal{C}_{g} \cup \{k\}$.

Let the MGs represent the vertices of a graph $G$ where an edge is drawn between a pair of vertices if the corresponding MGs have co-channel interference greater than a certain threshold ($\gamma_{th}$). Let  CU channel represent the color. Accordingly, the corresponding interference matrix $A_d$ is defined as follows:
\begin{align}
\label{deqn_ex1a}
A_d({g,j}) = 
\begin{cases}
    1, & \text{if } \mathcal{C}_g \cap \mathcal{C}_j = \emptyset, \\
    1, & \text{if } \mathcal{C}_g \cap \mathcal{C}_j \neq \emptyset \text{ and } | h_{g,r_j}  - h_{j,r_g} | < \gamma_{th}, \\
    0, & \text{otherwise},
\end{cases}
\end{align} 
where $g$ and $j$ represent any two MGs from the set $\mathcal{G}$.

The vertices of the interference graph $G$ are subjected to a coloring process for channel allocation, detailed in Algorithm~\ref{alg:Channel_assign}. The proposed algorithm iteratively updates the parameter $\gamma_{th}$ whenever the subroutine in Algorithm~\ref{alg:Graph_coloring} triggers a $flag$ indicating its failure. As these iterations progress, the value of $\gamma_{th}$ steadily decreases. Consequently, there exists a possibility that $\gamma_{th}$ could decrease to an extent where each individual node in the graph $G$ becomes isolated, devoid of any neighboring nodes. When this scenario is reached, each node in the graph effectively stands alone and independent, without any interference from neighboring nodes. This implies that the graph $G$ can be colored using a single color, and thus establishing the convergence of Algorithm~\ref{alg:Channel_assign}.

\begin{algorithm}[]
 \caption{Distributed Channel Allocation Scheme}
 \label{alg:Channel_assign}
 \begin{algorithmic}[1]
  \STATE \textbf{Input:} $G$, $\gamma_\text{th}$, $flag=0$, $\delta$, where $\delta$ denotes the small percentage change in $\gamma_\text{th}$
  \STATE \textbf{Return:} Channel allocation of each MG
  \STATE Color the graph $G$ according to Algorithm \ref{alg:Graph_coloring} and update the $flag$.
  
  \WHILE{$flag$ or $colors\_unique < N_{c}$}
  \IF{flag}
   \STATE Update $\gamma_\text{th} \leftarrow \gamma_\text{th} - \delta$.
  \ELSE 
      \STATE Update $\gamma_\text{th} \leftarrow \gamma_\text{th} + \delta$.
  \ENDIF
    \STATE Update $A_d$ according to Equation~\eqref{deqn_ex1a} and form a new graph $G$ again.
    \STATE Set the $flag=0$ and color the new graph $G$ according to Algorithm \ref{alg:Graph_coloring} and update the $flag$.
    
  \ENDWHILE 
\end{algorithmic}
\end{algorithm}

\begin{algorithm}[]
 \caption{Distributed Graph Coloring for a given $\gamma_\text{th}$}
 \label{alg:Graph_coloring}
 \begin{algorithmic}[1]
     \STATE \textbf{Input:}   $G$, $\mathcal{C}_{g}$ , $flag$, $N\_g$: the set of neighbors of node $g$, $\forall g \in \mathcal{G}$
     \STATE \textbf{Return:} The color assignment for each node, a $flag$ indicating whether the algorithm failed to assign color to each node, and total number of colors ($colors\_unique$).
    \WHILE{there exist uncolored node $g$ and non-empty set $\mathcal{C}_{g}$}
%
     \FORALL{unassigned nodes }
      \STATE Assign a color from $\mathcal{C}_{g}$ to $g$.
     \ENDFOR
     \STATE Each node informs its neighbors about its color.
           \IF{node $g$  and  $j \in N\_g$ have the same color}
       \STATE Unassign the color of the node with larger $\left |\mathcal{C}_{g} \right|$. 
      \ENDIF  
     \STATE Again each nodes will inform its neighbors about it assign color.
     \STATE Each unassigned node $g$ will update its $\mathcal{C}_{g}$ by removing the colors of its neighbors.
      
    \ENDWHILE
    \STATE If there is no unassign node then $flag =0$, else $flag =1$.
   
 \end{algorithmic}
\end{algorithm}

The proposed distributed channel allocation scheme effectively assigns channels to all D2D MGs which contribute positively to sum-throughput and ensures minimum co-channel interference among the MGs and CUs. The computation complexity of channel allocation Algorithm \ref{alg:Channel_assign} is dependent on the Algorithm \ref{alg:Graph_coloring} and equation (\ref{deqn_ex1a}). The computation complexities of graph coloring Algorithm \ref{alg:Graph_coloring} and equation (\ref{deqn_ex1a}) are $O(I_dN_{p}^{2})$ and $O(N_{p}^{2})$ respectively, where $I_d$ is the maximum number of iterations for Algorithm \ref{alg:Graph_coloring} to converge. The $\gamma_\text{th}$ updation step complexity is $O(1)$. Therefore, the overall computation complexity of the channel allocation is $O(I_cI_dN_{p}^{2})$, where $I_c$ is the maximum number of iterations required for Algorithm \ref{alg:Channel_assign} to converge.

\section{Distributed Power Allocation}
\label{power_allocation}
We propose a distributed power allocation scheme to address the uplink interference at the BS caused by MGs sharing the channel with CUs. The primary aim is to formulate a scheme that encourages MGs to adjust their power levels to minimize interference for the co-channel CUs. In this approach, each unassigned MG chooses a power within the permissible range, which is determined by the minimum power ($p_{min}$) and maximum power ($p_{max}$). To ensure that an MG causing more interference and operating with a higher power reconsiders its power selection, we adopt a conflict resolution approach that allows MGs that cause greater interference to the co-channel CUs to be assigned lower power levels than the MGs that cause less interference. The MGs iteratively change their power during conflict resolution while communicating this information with co-channel MGs. Upon the resolution of conflicts, if the overall interference caused by all co-channel MGs exceeds the predefined interference threshold ($\gamma^{k}_{CU}$), the next step is to examine the interference offered by each MG at the BS separately. For this, we employ a threshold value $\gamma^{k}_{MG}$, calculated by dividing the CU's interference threshold $\gamma^{k}_{CU}$ by the total number of MGs sharing the channel. If the interference generated by a specific MG exceeds $\gamma^{k}_{MG}$, its power allocation is revoked. Furthermore, the minimum power ($p_{min}$) and maximum power ($p_{max}$) values are adjusted slightly downward, and the whole process repeats. The detailed steps are shown in Algorithm~\ref{alg:power_allocation}. 

This iterative process continues until all MGs have power that permits them to transmit without violating the BS's interference threshold. This ensures convergence of the Algorithm~\ref{alg:power_allocation}. Furthermore, when while addressing the MGs with severe interference at the BS, variations in power may eventually pull down the power levels of these interfering MGs to zero. As a result, these MGs will satisfy the interference threshold at the BS, proving the convergence of Algorithm~\ref{alg:power_allocation}. 

\begin{algorithm}[]
\caption{Distributed Power Allocation Scheme}
\label{alg:power_allocation}
\begin{algorithmic}[1]
    \STATE \textbf{Input:} Set of unassigned MGs, $\gamma^{k}_{CU}$, $p_{min} = P_g^{max} - \beta $ and $p_{max} = P_g^{max}$, where $\beta$ denotes the small change.
    \STATE \textbf{Return:} Power allocation for each MG.
    
    \STATE Assign random power levels within $[p_{min}, p_{max}]$ to each unassigned MG.
    \STATE Resolve conflicts: Adjust power levels distributively to ensure MGs causing more interference allocate lower power levels than those causing less interference.
   
    \IF{interference caused by all MGs sharing channel $k$ exceeds $\gamma^{k}_{CU}$}
    \STATE Calculate $\gamma^{k}_{MG} = \frac{\gamma^{k}_{CU}}{\mathbb{G}_{k}}$.
        \FOR{each MG}
            
            \IF{MG's interference is greater than $\gamma^{k}_{MG}$}
                \STATE Unassign power of that MG.
               
            \ENDIF
        \ENDFOR
         \STATE Update $p_{min} \leftarrow p_{min} - \beta$ and $p_{max} \leftarrow p_{max} - \beta$.
        \STATE Repeat the power allocation process.
    \ENDIF
\end{algorithmic}
\end{algorithm}

The computation complexity of Algorithm \ref{alg:power_allocation} depends mainly on number of iterations ($I_p$) for the algorithm to converge and conflict resolution step (line 4). The conflict resolution step has the worst case complexity of $O(N_{p}^2)$. Therefore, overall complexity of proposed power algorithm is $O(I_pN_{p}^2)$.

\section{Performance Evaluation}
\label{simulation}
To show efficacy of proposed scheme, numerical simulations are performed to compute the total system sum-throughput, while varying different parameters. In the context of channel allocation, we compare our proposed channel allocation scheme with Interference-Aware Channel Allocation (IACA)~\cite{bhardwaj2017interference} and Random Channel Allocation (RCA) schemes but not with distributed scheme in \cite{elnourani2021distributed}, as it has already been outperform by IACA in terms of sum-throughput \cite{elnourani2021distributed}. While for power allocation, we compare our proposed power allocation scheme with water-filling power allocation (WFPA)~\cite{bansal2008optimal}, STIM power allocation (STIMPA)~\cite{bhardwaj2017interference} and equal power allocation (EPA). To ensure a fair and relevant comparison while comparing different channel allocation schemes, we pair all channel allocation schemes with our proposed power allocation approach. Similarly, while comparing power allocation schemes, we pair all power allocation schemes with our proposed channel allocation scheme. Simulations parameters are listed in Table~\ref{table_1}. Each data point in the following plots is an average of $500$ network instances.
\begin{table}[!t]
\caption{SIMULATION PARAMETERS}
\centering
\begin{tabular}{|c||c|}
\hline
Cell Radius & 500 meters\\
\hline
Noise Power & -114 dBm\\
\hline
Path Loss Exponent & 3.6\\ 
\hline
Shadowing & 8 dB\\
\hline
Bandwidth & $10^6$ Hz\\
\hline

$P^{max}_{c},P^{max}_{g}$ & 30 dBm , 25 dBm\\
\hline
\end{tabular}
\label{table_1}
\end{table}


\begin{figure}[!t]
\centering
\includegraphics[width=\textwidth]{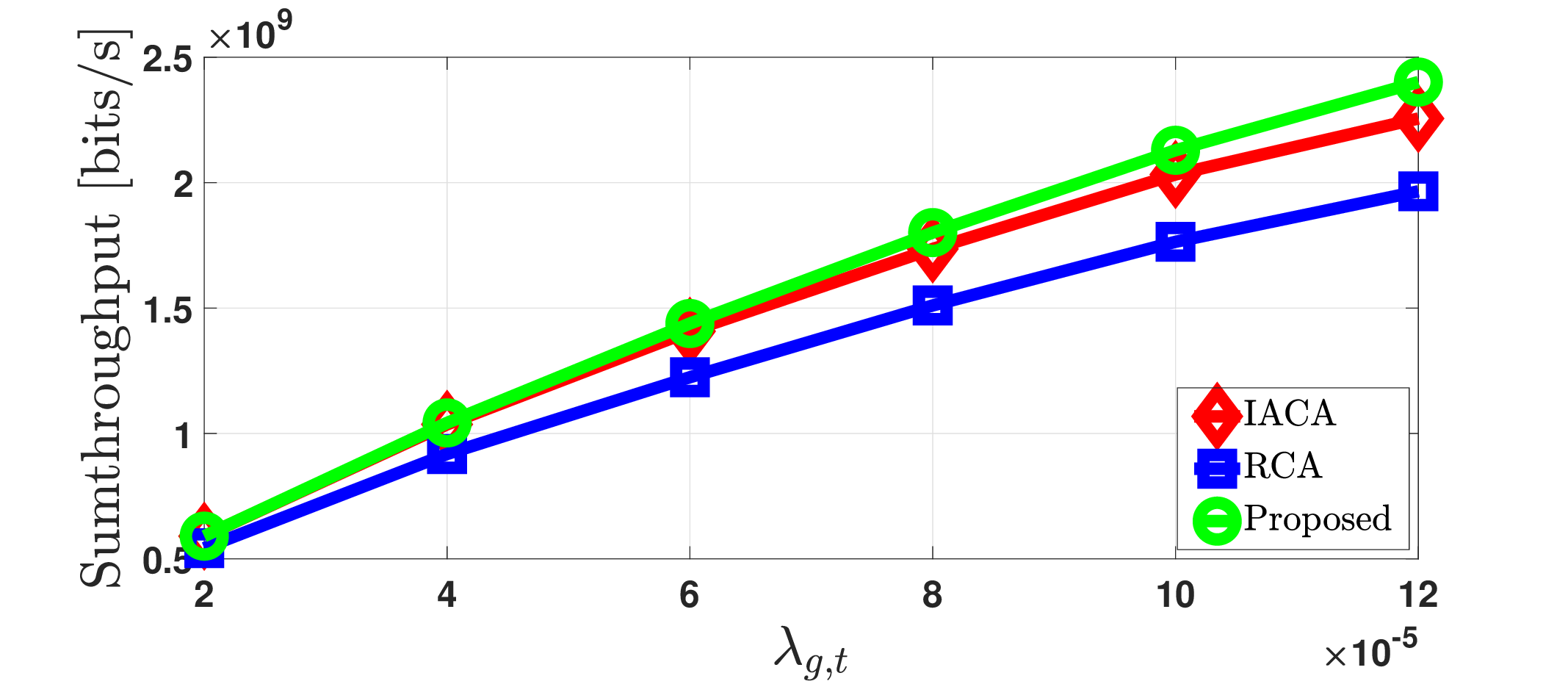}
\caption{Sum-throughput versus MGTx density for different channel allocation schemes.}
\label{fig_1}
\end{figure}

Fig.~\ref{fig_1} shows the comparison between the total sum-throughput vs the MG density ($\lambda{\textsubscript{g,t}}$) for different channel allocation algorithms. As the $\lambda{\textsubscript{g,t}}$ increases, the total sum-throughput also increases. As $\lambda_{g,t}$ doubles, the rate of increase in the sum-throughput decreases from 76.5\% to 38\%, and further decreases to 13\% for the proposed scheme. This decline in the rate of increase is attributed to the co-channel interference that arises due to the growing number of MGs. Notably, our proposed scheme outperforms both the IACA and the RCA. With each doubling of $\lambda_{g,t}$, the increase in sum-throughput for our proposed scheme consistently surpasses the IACA by approximately around 0.1-6\% and the RCA by approximately 8-22\%. This observation indicates that our proposed distributed channel scheme effectively mitigates co-channel interference, outperforming IACA and RCA. 

\begin{figure}[!t]
\centering
\includegraphics[width=\textwidth]{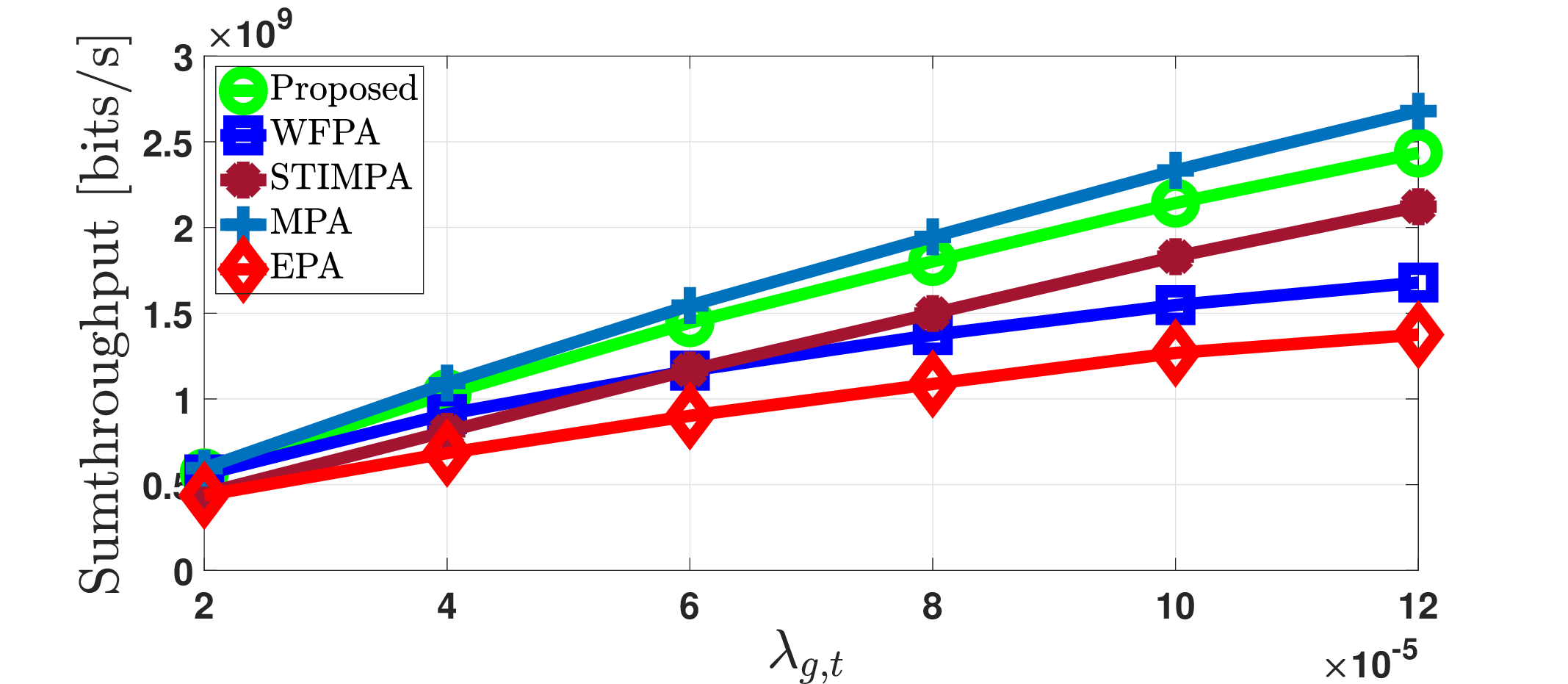}
\caption{Sum-throughput versus MGTx density for different power allocation schemes.}
\label{fig_2}
\end{figure}

Fig.~\ref{fig_2} shows the comparison between the total sum-throughput vs the MGs density ($\lambda{\textsubscript{g,t}}$) for different power allocation algorithms. Initially, the sum-throughput of the proposed scheme is 1\%, 27\% and 30\% greater than WFPA, STIMPA and EPA, respectively. Furthermore, as $\lambda_{g,t}$ doubles, the sum-throughput of the proposed scheme consistently outperforms the aforementioned power allocation schemes by increasingly significant margins.
These results indicate that the proposed distributed power allocation scheme accommodates co-channel interference in a better manner than the existing schemes.
Although, Fig. \ref{fig_2} shows that MPA performs better than the proposed scheme but it does not respect the CU threshold, and is thus impractical.

\begin{figure}[!t]
\centering
\includegraphics[width=\textwidth]{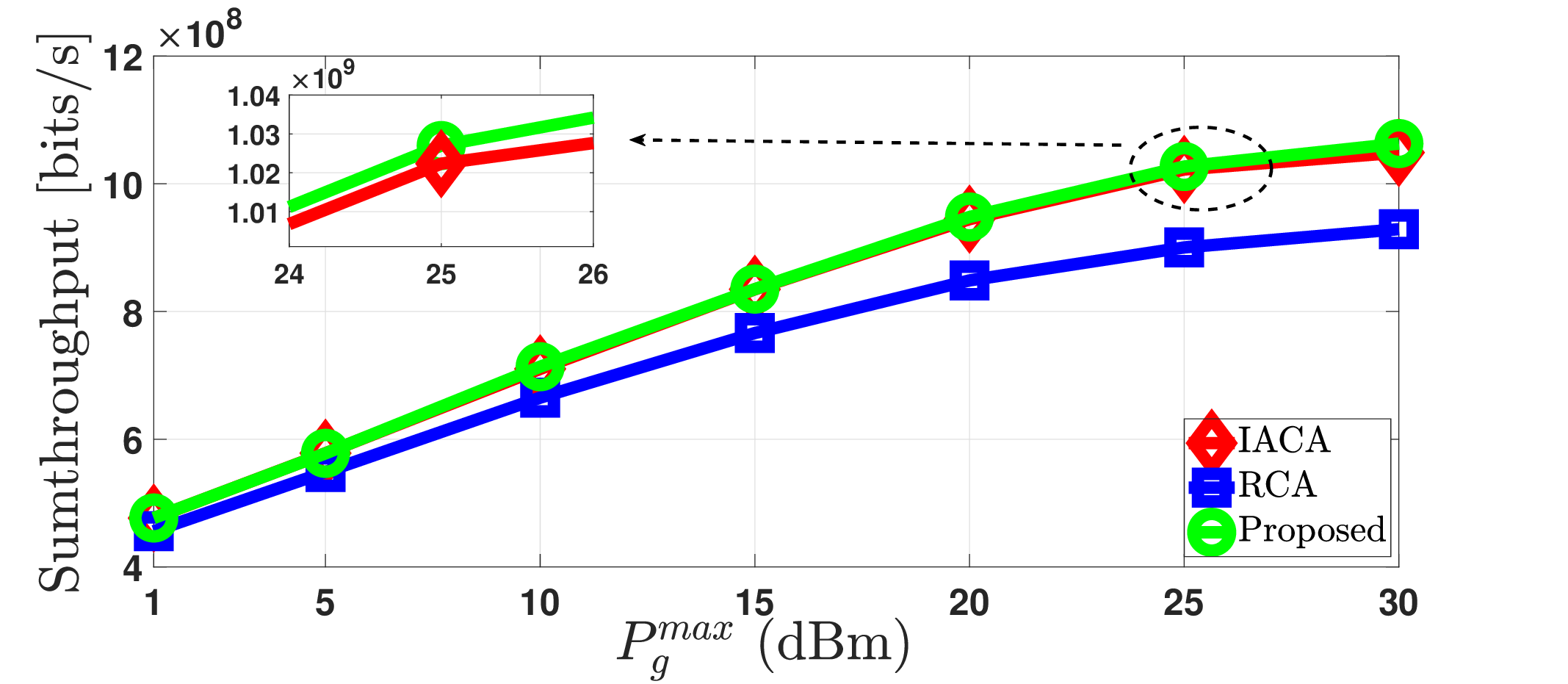}
\caption{Sum-throughput versus $P^{max}_{g}$  for different channel allocation schemes.}
\label{fig_3}
\end{figure}

Fig.~\ref{fig_3} shows the variation of sum-throughput with the variation in $P^{max}_{g}$ while keeping the \textit{d\textsubscript{r}} fixed for different channel allocation schemes. As the $P^{max}_{g}$ increases, the sum-throughput also increases. Initially, when $P^{max}_{g}$ is increased from 1 dBm to 5 dBm and then to 10 dBm, the sum-throughput for proposed scheme experiences significant increases of 21\% and 23.5\% respectively. However, as $P^{max}_{g}$ is further increased to 30 dBm, the subsequent increase in sum-throughput diminish to 3.5\%. This diminishing trend in the rate of increase of sum-throughput can be attributed to the concurrent rise in co-channel interference as the $P^{max}_{g}$ value increases. As the $P^{max}_{g}$ reaches higher levels, the interference caused by neighboring MGs becomes stronger. The results clearly demonstrate the superiority of the proposed channel allocation scheme over both the IACA and RCA schemes, as depicted in Fig.~\ref{fig_3}. While the performances of the IACA and proposed schemes are comparable for most cases, it is noteworthy that the proposed scheme outperforms IACA scheme specifically for high values of $P^{max}_{g}$. At these higher power levels, the impact of co-channel interference becomes more pronounced, and it is observed that the IACA is more adversely affected compared to the proposed scheme. Consequently, the sum-throughput of the proposed scheme surpasses that of the IACA by approximately 0.5\% to 1.5\%.

\begin{figure}[!t]
\centering
\includegraphics[width=\textwidth]{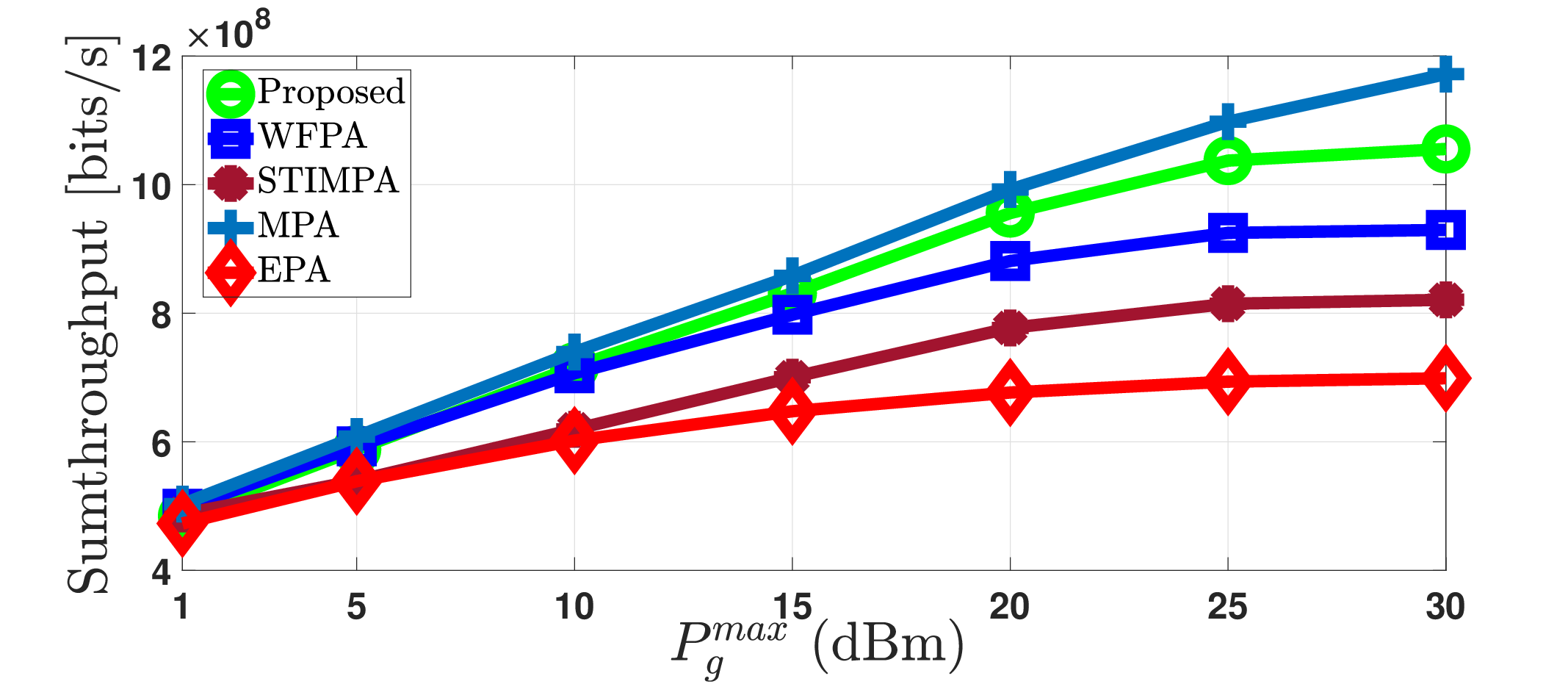}
\caption{Sum-throughput versus $P^{max}_{g}$ for different power allocation schemes.}
\label{fig_4}
\end{figure}
 
In Fig.~\ref{fig_4}, the variation of sum-throughput with changes in $P^{max}_{g}$ is illustrated while keeping the distance $d_r$ fixed for different power allocation schemes. The results highlight the performance differences among these schemes in the presence of co-channel interference. For lower power levels, where severe co-channel interference is not a significant factor, the WFPA exhibits a slightly higher sum-throughput, approximately 1-2\%, compared to the proposed scheme. However, as $P^{max}_{g}$ increases beyond 10 dBm, co-channel interference starts to play a substantial role in system performance. For every 5 dBm increase in $P^{max}_{g}$, the proposed scheme consistently outperforms the WFPA in terms of sum-throughput.
For the initial power level of 1 dBm, the sum-throughput of the STIMPA is marginally higher by approximately 0.5\% compared to the proposed scheme. However, as $P^{max}_{g}$ increases, the performance gap between the proposed scheme and the STIMPA widens significantly.
These findings clearly indicate that at higher power levels, when co-channel interference becomes more severe, the proposed scheme demonstrates superior performance by effectively handling and mitigating the impacts of co-channel interference.

\textit{Discussion:} The proposed scheme outperforms existing ones in terms of the system sum-throughput. It also provides a range of advantages. Firstly, the distributed approach has greater scalability as each node only needs to communicate with its immediate neighbors, allowing it to handle larger networks without incurring excessive communication and computational overhead. Secondly, the distributed approach can provide lower latency due to its decentralized nature, where each node makes its own channel and power allocation decisions based on information of neighboring nodes. This enables faster and more efficient decision-making, which may improve network performance. Finally, the distributed approach is more fault-tolerant compared to the centralized scheme since each node operates independently, and the failure of a single node does not disrupt the entire network.

\section{Conclusion and Future Work}
\label{conclusion}
The results in this work suggest that distributed channel and power allocation can enhance the performance of underlay D2D multicast networks while having little impact on cellular users' performance. Apart from better sum-throughput, distributed schemes have several advantages due to their decentralized nature, including better scalability, fault tolerance, and lower latency. Our results emphasize the potential of distributed schemes for resource allocation in underlay D2D multicast networks, as well as the need for additional research in this area, specifically in the scenarios where multicast groups may have different priorities, and the resource allocation among the multicast groups is subjected to some fairness criteria.

\end{document}